# Gate-Tunable Multi-Band van der Waals Photodetector and Polarization Sensor

Daozhi Shen[1,2,3], HeeBong Yang[2,4], Tarun Patel[2,5], Daniel A. Rhodes[6], Thomas Timusk[7], Y. Norman Zhou[8,9], Na Young Kim[2,4,9], and Adam W. Tsen[2,3,9]*

[1] School of Mechanical Engineering, Shanghai Jiao Tong University, Shanghai, 200240, China

[2] Institute for Quantum Computing, University of Waterloo, Waterloo, ON, Canada N2L 3G1

[3] Department of Chemistry, University of Waterloo, Waterloo, ON, Canada N2L 3G1

[4] Department of Electrical and Computer Engineering, University of Waterloo, Waterloo, ON, Canada N2L 3G1

[5] Department of Physics and Astronomy, University of Waterloo, Waterloo, ON, Canada N2L 3G1

[6] Department of Materials Science and Engineering, University of Wisconsin-Madison, WI, USA

[7] Department of Physics and Astronomy, McMaster University, Hamilton, Ontario, Canada L8S 4M1

[8] Centre for Advanced Materials Joining and Department of Mechanical and Mechatronics Engineering, University of Waterloo, Waterloo, Ontario, Canada N2L 3G1

[9] Waterloo Institute for Nanotechnology, University of Waterloo, Waterloo, Ontario, Canada N2L 3G1

* Corresponding author: awtsen@uwaterloo.ca

**Abstract: A single photodetector with tunable detection wavelengths and polarization sensitivity can potentially be harnessed for diverse optical applications ranging from imaging and sensing to telecommunications. Such a device will require the combination of multiple material systems with different structures, bandgaps, and photoelectrical responses, which is extremely difficult to engineer using traditional epitaxial films. Here, we develop a multi-functional and high-performance photosensor using all van der Waals materials. The device features a gate-tunable spectral response that is switchable between near-infrared/visible and short-/mid-wave infrared, as well as broadband operation, at room temperature. The linear polarization sensitivity in the telecommunications O-band can also be directly modulated between horizontal, vertical, and nonpolarizing modes. These effects originate from the balance of photocurrent generation in two of the active layers that can be manipulated by an electric field. The photodetector features high detectivity ($>10^9$ cmHz$^{1/2}$W$^{-1}$) together with fast operation speed (~ 1 MHz) and can be further exploited for dual visible and infrared imaging.**

## Introduction

Many technology-intensive fields, such as healthcare[1], materials characterization[2], remote sensing[3], and optical communications[4], rely on photosensitive devices that can detect light intensity or polarization within critical spectral bands (e.g. mid-wave infrared (MWIR), short-wave infrared (SWIR), near-infrared (NIR), visible (Vis), ultraviolet (UV), etc.). While in principle this can be achieved using multiple sensor units and optical components, a single photodetector with an electrically tunable spectral and/or polarization response can be a simple and powerful alternative, finding ubiquitous use across many disciplines. Although multi-color photodetectors have been successfully implemented using mature technologies such as quantum wells and epitaxial superlattices[5,6], they suffer from expensive heteroepitaxy growth methods, limited materials selection due to lattice-matching requirements, and the need to operate at low temperatures to avoid thermal noise[7]. Furthermore, their polarization sensitivity usually relies on the use of external optical components, adding to the complexity of the system.

Recently, two-dimensional (2D) materials have shown great promise for multi-color photodetection, as the absence of surface dangling bonds allows for the integration of 2D materials into complex heterostructures regardless of lattice mismatch[8,9]. This offers a large degree of freedom for configuring devices with different structures and functions. Many 2D materials and their heterojunctions, including twisted graphene (Gr)[10], black phosphorus (BP)[11, 12], $MoTe_2$[13], $MoS_2$[14], and $PdSe_2$[15] have been used to build photodetectors with high detectivity, fast speed, broadband sensitivity, and room temperature operationality. So far, very few studies have demonstrated electrical tuning of the spectral response for such devices, while simultaneous control over polarization sensitivity has never been achieved.

In this work, we describe a gate-tunable van der Waals (vdW) photodetector incorporating a vertical Gr / 2H-$MoTe_2$ / BP heterostructure as the photoactive and/or charge collection layers. Previously, we have shown such a trilayer structure can be sensitive to radiation across a decade in energy from the MWIR to UV, or 0.3 to 5 eV, for optimal voltage bias at room temperature[16]. With the addition of a Gr gate and a hexagonal boron nitride (hBN) dielectric layer, the detector can now be robustly tuned to operate only in the MWIR and SWIR bands (0.3 − 0.9 eV) or NIR and Vis bands (0.9 − 2.5 eV), as well as in broadband mode (0.3 − 2.5 eV), all under zero-bias conditions. This effect is possible as the flatband condition for quenching the photoresponse in either the $MoTe_2$ or BP layers is achieved under different gate fields, which the top Gr contact does not fully screen. At the same time, in the telecommunications O-band (0.91 − 0.98 eV) where both BP and $MoTe_2$ contribute to the photocurrent generation, the polarization sensitivity for the detector can also be continuously and electrically tuned. Finally, we show how our detector can be used for dual-band imaging of micron scale objects in a confocal geometry. Our results demonstrate the power of using designer vdW heterostructures for next-generation multi-functional optoelectronics.

**Results and Discussion**

Figure 1a shows a schematic of our device geometry and measurement circuit together with an optical image of the active area. Few-layer Gr provides a transparent top gate and hBN acts as both a dielectric and protective layer for the reactive BP underneath. The middle Gr layer provides a transparent top contact with low carrier concentration to support electrostatic gating of the underlying layers. Their thicknesses are shown in Figure S1. As can be seen from the energy diagram in Figure 1b, 2H-$MoTe_2$ absorbs light energy above 0.9 eV [16-18], while BP serves both as a bottom conductor and absorber for light energy down to its 0.3 eV bandgap[16, 19–21]. The heterostructure was assembled by dry-transfer[22] within a nitrogen-filled glovebox and placed on an insulating $Al_2O_3$ substrate with pre-patterned Au/Ti electrodes. Figure 1c shows a cross-sectional transmission electron microscopy image of the same heterostructure taken after device measurements, from which the high-quality atomically abrupt interfaces can be clearly resolved.

To begin characterization and analysis of the transport properties of our photodetectors, we first measure the conductance at low bias with changing gate voltage $V_G$. In the inset of Figure 1d, we see that the conductance can be tuned by over two orders of magnitude. As the junction resistance is dominated by the $MoTe_2$, this indicates that its Fermi level can still be strongly modulated despite the presence of the Gr contact above [23,24]. We note, however, that the on/off ratio is smaller compared to that of lateral $MoTe_2$ transistors without a graphene screening layer (see Figure S2). For $V_G \geq 0$ V, the device exhibits diode-like rectifying behavior, as can be seen from the gate-dependent *I-V* characteristics shown in the main panel of Figure 1d (see Figure S3 for semilog plot of *I-V*). For $V_G < 0$ V, the device becomes more conductive for both bias directions and the *I-V* curve becomes symmetric.

We have performed complete finite-element simulations in COMSOL to model this transport behavior. The gated-dependent parameters and Fermi levels for $MoTe_2$ and BP are summarized in Table S1 and Figure S4. In general, there are built-in electric fields across the $MoTe_2$ and interfacial BP region because of the energy mismatch between the materials (see Figure 1b). With an applied gate, the fields can be substantially modified, with different flat-band conditions for the BP and $MoTe_2$ layers as we shall discuss in the context of the photoresponse. The simulated gate-dependent *I-V* characteristics are plotted in Figure 1e and show good qualitative agreement with the experimental data.

We now discuss the gate-dependent photoresponse under zero-bias conditions (which minimizes the dark current). Figure 2a shows photocurrent ($I_{pc}$) vs. gate voltage $V_G$ under illumination with two different laser wavelengths: λ = 658 nm ($MoTe_2$ dominant response) and 1550 nm (BP dominant response). We have used relatively low laser power levels to remain within the linear regime (see Figure S5). While the 658 nm photocurrent response shows ambipolar behavior (with a zero-level crossing at $V_G$ = 2.7 V), photocurrent for λ = 1550 nm is strongly activated only for positive $V_G$. This feature is consistent with results from the full illuminated *I-V* characteristics shown in Figure S6. Furthermore, there is a region of gating (−2 V ≲ $V_G$ < 2.7 V) where the sign of the two photocurrent signals is opposite, suggesting opposite band-bending slopes in the $MoTe_2$ and BP layers. To confirm that these effects are spatially uniform across the Gr / $MoTe_2$/ BP junction, we performed scanning photocurrent microscopy (see Figure 2b) using the same two wavelengths and specially chosen gate voltages ($V_G$ = −1 V and $V_G$ = 5 V). First, we can clearly see that the active region is well-localized within the overlap area between the Gr contact, $MoTe_2$, and BP, regardless of $V_G$ or λ. For λ = 658 nm, the two gate voltages yield photocurrent signals of different sign, whereas for λ = 1550 nm, photocurrent is observed for $V_G$ = 5 V only. These results are consistent with the measurements shown in Figure 2a for fixed laser position.

Next, we characterize the full gate-dependent photocurrent spectral response of our detector from the MWIR (0.25 eV) to Vis (2.5 eV) using Fourier transform infrared spectroscopy. The response at higher energies is difficult to assess using this technique due to the low intensity of the tungsten light source. The results are shown in Figure 2c as a normalized 2D false-color plot, which can be vertically subdivided into four regions marked by the horizontal dashed lines and defined by their spectral response as the MWIR/SWIR band, NIR/Vis band, mixed band, and broadband. A representative photocurrent spectrum for each region is overlaid in gray. In the NIR/Vis band ($V_G$ < −2 V), the device only shows a photoresponse for energies above 0.9 eV, with a strong peak at ~ 1.1 eV. The latter coincides with the A exciton of $MoTe_2$ [16,18], indicating that only $MoTe_2$ contributes to the photocurrent signal in this gate region. In the MWIR/SWIR band (2 V < $V_G$ < 3 V), the device shows strong photoresponse between 0.3 eV and 0.9 eV, and the A exciton peak is absent. This indicates that only BP is photoactive here. In the broadband region ($V_G$ > 3 V), the device exhibits a continuous response from 0.3 eV to over 2.0 eV as photocarriers from both BP and $MoTe_2$ contribute. It should be noted that while such broadband photodetection can be achieved by applying a proper bias voltage[16], accessing each band individually can only be realized by gate tuning

(see Figure S7). Finally, in the mixed band ($-2$ V $< V_G <$ 2 V), a broad spectral response can also be achieved; however, a node in the spectrum is seen at ~ 0.95 eV due to a canceling of opposite photocurrent signals from the BP and MoTe$_2$ layers in this gate window.

The changes observed in the photocurrent spectrum with gating can be understood from our COMSOL model. In Figure 2d, we show simulated band structures for each of the gate regions discussed above to understand the behavior under illumination. For $V_G < -2$ V, the bands across the entire BP layer are flat, and so only MoTe$_2$ will contribute to the photoresponse at excitations greater than ~ 0.9 eV. Importantly, the direction of the MoTe$_2$ photocurrent is positive according to our measurement circuit. As $V_G$ is increased in the positive direction, the bands on the right are pulled down in energy. This creates band bending at the interfacial BP region with different slope than across the MoTe$_2$. In the region $-2$ V $< V_G <$ 2 V, the low energy photocurrent generated in BP is therefore in the opposite direction to the high energy photocurrent generated in MoTe$_2$, yielding a node at the precise energy and gate voltage where the two cancel. For $2 < V_G < 3$ V, the MoTe$_2$ bands become flat, and so its photoresponse becomes quenched, leaving only the BP infrared photoactive. Finally, upon increasing the gate further ($V_G > 3$ V), the MoTe$_2$ bands bend in the opposite direction (with the same slope as that for BP), allowing both materials to contribute photocurrent of the same sign.

The high degree of spectral tunability is the primary advantage of our photodetector concept. It relies on the special band interface between BP and MoTe$_2$ together with the low carrier density of the Gr contact to allow for the penetration of the gate field deep into the vertical heterostructure. We have further characterized the key performance metrics of responsivity ($\mathfrak{R}$), external quantum efficiency (EQE), and specific detectivity ($D^*$) all as a function excitation energy and $V_G$, and obtain peak values of 0.17 A/W, 17%, and $2\times10^9$ cmHz$^{1/2}$W$^{-1}$, respectively (see Figures S8 and S10). We have also measured a $-3$ dB electrical bandwidth of $0.92-0.95$ MHz for both infrared and visible illumination (see Figure S11), while the spectral crosstalk between the MWIR/SWIR and NIR/Vis bands is $5-7$% (see Figure S12). Overall, these results demonstrate that our detectors perform on par commercial (non-amplified and uncooled) photodetectors with fixed spectral response[25].

At the same time, the competing photocurrent signals from the BP and MoTe$_2$ in the mixed region provides an opportunity to manipulate the polarization sensitivity of our detector, as the anisotropic structure of BP yields stronger (weaker) photocurrent for polarization along the armchair (zigzag) direction [11,26]. In Figures 3a and 3b, we first show the gate-dependent linear polarization angle dependence of $|I_{pc}|$ for $\lambda = 1550$ nm and 658 nm, where the photocurrent is dominated by BP and MoTe$_2$, respectively. At all applied gate voltages, the former exhibits a dipole-like structure, while the latter shows little angle dependence—both are the expected photocurrent responses for those two materials individually. We next choose $\lambda = 1310$ nm, or 0.95 eV, which is in the center of the telecommunications O-band and where both BP and MoTe$_2$ show substantial photocurrent. Here, the polarization angle dependence can be strongly modulated with the gate, as shown in Figure 3c. At $V_G = -5$ V (NIR/Vis band), only the MoTe$_2$ is photoactive, and so we observe a circular pattern (like that in Figure 3b). At $V_G = 3$ V (MWIR/SWIR band) where only BP responds, we see a dipole pattern where the maxima are directed along the 0 and 180 degrees (as with Figure 3a). At intermediate gate voltages in the mixed band, the photocurrent generated from BP subtracts from that of MoTe$_2$, and so the maximum lobes are now directed along 90 and 270 degrees. The polarization sensitivity can thus be switched between unpolarized response and horizontal and vertical linear polarization at this wavelength.

Figure 3d shows a 2D false-color map of the original photocurrent data from which the circular angle plots in Figure 3c are derived. One can clearly see that the photocurrent anisotropy and polarization angles for which $|I_{pc}|$ is maximum changes as is $V_G$ changed. We have also calculated the photocurrent anisotropy (defined as $|I_{pc}^{90°}|/|I_{pc}^{0°}|$) as a function of gate voltage explicitly, and the results are shown in Figure 3e. We can tune the anisotropy over nearly two orders of magnitude; no other photodetector has demonstrated such striking polarization control as far as we are aware. A comparison of our device with reported multiband or dual-band or multiband detectors based on 2D materials is summarized in Table S1.

Finally, we demonstrate experimentally that our photodetector can be used effectively for dual-band imaging in a confocal geometry. Our setup is depicted in Figure 4a. A broadband supercontinuum laser is focused and raster-scanned across our test samples and the reflected beam is focused on our detector. Figure 4b shows confocal images taken of two different material systems and with the detector biased at $V_G = -4$ V (NIR/Vis band) and $V_G = 2.5$ V (MWIR/SWIR band). The first system (upper panels) consists of nearby BP and $WS_2$ (bandgap ~ 1.4 eV) flakes transferred on a sapphire substrate. When the detector is operated in the NIR/Vis band, both flakes can be seen by the detector as the excitation energy of NIR/Vis photons from supercontinuum laser exceeds the bandgap for both semiconductors and becomes partially absorbed (see upper left panel in Figure 4b). In the MWIR/SWIR band, only the BP flake is visible to the detector as $WS_2$ becomes fully transparent to MWIR/SWIR photons (see upper right panel in Figure 4b). The second material system (lower panels) is an InGaAs wafer with local microdefects (scratches, dust, etc.). When imaged in the MWIR/SWIR band, several line features can be observed that are absent for NIR/Vis imaging. They likely correspond to step flow stripes caused by the accumulation of strain[27].

**Conclusion and outlook**

These proof-of-principle results demonstrate the power of our tunable detector in resolving objects under different parts of the electromagnetic spectrum without any change in external optics. While the current design already provides high performance across an extremely large spectral range together with room temperature operability, the concept can be easily extended to other 2D semiconductors with various bandgaps to target other niche spectral windows for robust gate and polarization control. Further enhancement of the device responsivity and detectivity may potentially be achieved by incorporating on-chip waveguides[28] and plasmonic structures[29] as well as thermoelectric cooling for lower temperature operation.

**Materials and Experimental Methods**

**Crystal synthesis**

Black phosphorus (BP), hexagonal boron nitride (hBN), and graphite single crystals were purchased from HQ graphene. $2H-MoTe_2$ single crystals were synthesized by a self-flux method using Te as flux. Mo powder (99.9975%, Alfa Aesar 12972) and Te broken ingot (99.9999+%, Alfa Aesar 12607) were loaded into an alumina Canfield crucible (LSP Industrial Ceramics) and sealed in a quartz ampoule under vacuum (~$10^{-6}$ Torr). Subsequently, the ampoules were heated to 1100 °C over 11 h, held there for 3 days, and cooled to 500 °C at a rate of 0.9 °C/h. At 500 °C, the samples were centrifuged to separate the Te flux from the 2H-MoTe$_2$ crystals and quenched in air. The resulting 2H-MoTe$_2$ crystals were then removed from the ampoule, resealed in another quartz ampoule, and annealed in a temperature gradient ($T_{hot}$ = 415 °C, $T_{cold}$ = 100 °C) with crystals on the hot end for two days to remove any remaining excess Te.

**Device fabrication**

Gr / hBN/ Gr / 2H-MoTe$_2$ / BP heterostructures were fabricated using a layer-by-layer dry transfer method within a nitrogen-filled glove box (O$_2$ and H$_2$O levels below 0.1 ppm) to prevent oxidation. All thin flakes were first exfoliated from bulk crystals onto a Si substrate with a 300-nm-thick oxide layer. A stamp made of polydimethylsiloxane (PDMS) coated with a polycarbonate (PC) film was used to pick up the Gr, hBN, Gr, MoTe$_2$, and BP flakes from the Si substrate in sequence at 90 ºC. The entire stack was then transferred onto a clean sapphire substrate with pre-patterned Au (45 nm) / Ti (5 nm) electrodes at 120 ºC to delaminate the PC film and heterostructure from the PDMS. The residual PC was dissolved in chloroform at room temperature followed by rinsing in isopropyl alcohol. The thickness of the individual layers was subsequently determined using an atomic force microscope.

**Simulation**

The simulations of both the band diagrams and dark current-voltage (*I-V*) characteristics were performed using the COMSOL Multiphysics Semiconductor Module, similar to our previous work[15], from which the basic device parameters (effective density of states and doping concentration) were obtained for the case of $V_G$ = 0 in this work. The $V_G$ gating effect is modeled by changing the doping concentrations in the 2H-MoTe$_2$ and BP layers. In the vertical device structure, the top graphene contact does not fully screen the gate field. For the COMSOL simulation, the gate-dependent Fermi level $E_F$ is determined individually for the 2H-MoTe$_2$ and BP layers by adjusting the hole doping ($N_a$) and electron doping ($N_d$) concentrations for each. We then determine the overall energy band structures by combining the two layers together for a given gate condition.

**Characterization**

As prepared devices were wire-bonded onto chip carriers and then mounted into an optical cryostat (Montana Instruments) with a CaF$_2$ window and kept under vacuum at room temperature. Laser diodes with wavelengths in the visible (λ = 658 nm, Thorlabs) and infrared (λ = 1310 and 1550 nm, Thorlabs) were focused on the detectors using a 15× reflective objective (Beck Optronic Solutions, NA = 0.5). Bias and gate voltages were applied to the detector using a Source Measure Unit (Keithley 2450 SMU) and the current was measured using a pre-amplifier (Stanford Research Systems SR570). Photocurrent spectra of the detectors were characterized using a Bruker IFS 66v/S Fourier transform infrared spectrometer (FTIR). We measured the frequency response of our detector to obtain the −3 dB electrical bandwidth. The intensity of 658 nm (MoTe$_2$ response) and 1550 nm (BP response) wavelength laser diodes was modulated using a sinusoidal waveform with frequency ranging from 50 kHz to 2 MHz by a function generator (Stanford Research Systems DS345). For confocal imaging using van der Waals detector, a supercontinuum laser (YSL Photonics) with 420 nm to 2400 nm wavelength range was focused onto test samples (InGaAs wafer or Si substrate with 2D materials such as BP and WS$_2$, etc.) using a 15× reflective objective. The scanning mirror galvo in the beam path was used to raster the focused laser spot, while the reflected light was collected and focused onto our van der Waals detector. Gate voltages were applied to the device using the Keithley Source Measure Unit and the photocurrent was measured using the SR570 current pre-amplifier. A computer was used to control the beam position and to collect the photocurrent of the device under the irradiation of reflected beam from the test sample. The beam position and photocurrent information was used to obtain confocal reflection images.

**Supporting Information**

Cross-sectional transmission electron microscopy; energy dispersive spectroscopy; transport characteristics of MoTe$_2$ field-effect transistor; semilog *I-V* plot in dark; gate-dependent device modeling parameters; photocurrent $I_{pc}$ as function of laser power and gate voltage; full *I-V* characteristics under illumination; continuous 2D plots of photocurrent spectra vs light energy; photodetector performance metrics; spectral noise density of the device; *D*\* of photodetector under varies gate voltages at zero bias; frequency response.

**Acknowledgements**

A.W.T. acknowledges support from the National Science and Engineering Research Council of Canada (NSERC) and U.S. Army Research Office (W911NF-21-2-0136). D.S. acknowledges support from Shanghai Pujiang Program (22PJ1407600), the National Natural Science Foundation of China (52305388, BE0200030), and Shanghai Jiao Tong University Initiative Scientific Research Program (WH220402021). This research was undertaken thanks in part to funding from the Canada First Research Excellence Fund. H.B.Y. and N.Y.K. acknowledge the support of Industry Canada and the Ontario Ministry of Research & Innovation through Early Researcher Awards (RE09-068). H.B.Y. and N.Y.K. acknowledge CMC Microsystems for COMSOL Multiphysics. Y.N.Z. acknowledges NSERC and the Canada Research Chairs Program. T.T. was supported by NSERC grant "Optics of quantum materials." D.A.R. acknowledges the support provided by the University of Wisconsin-Madison Office of the Vice Chancellor for Research and Graduate Education with funding from the Wisconsin Alumni Research Foundation.

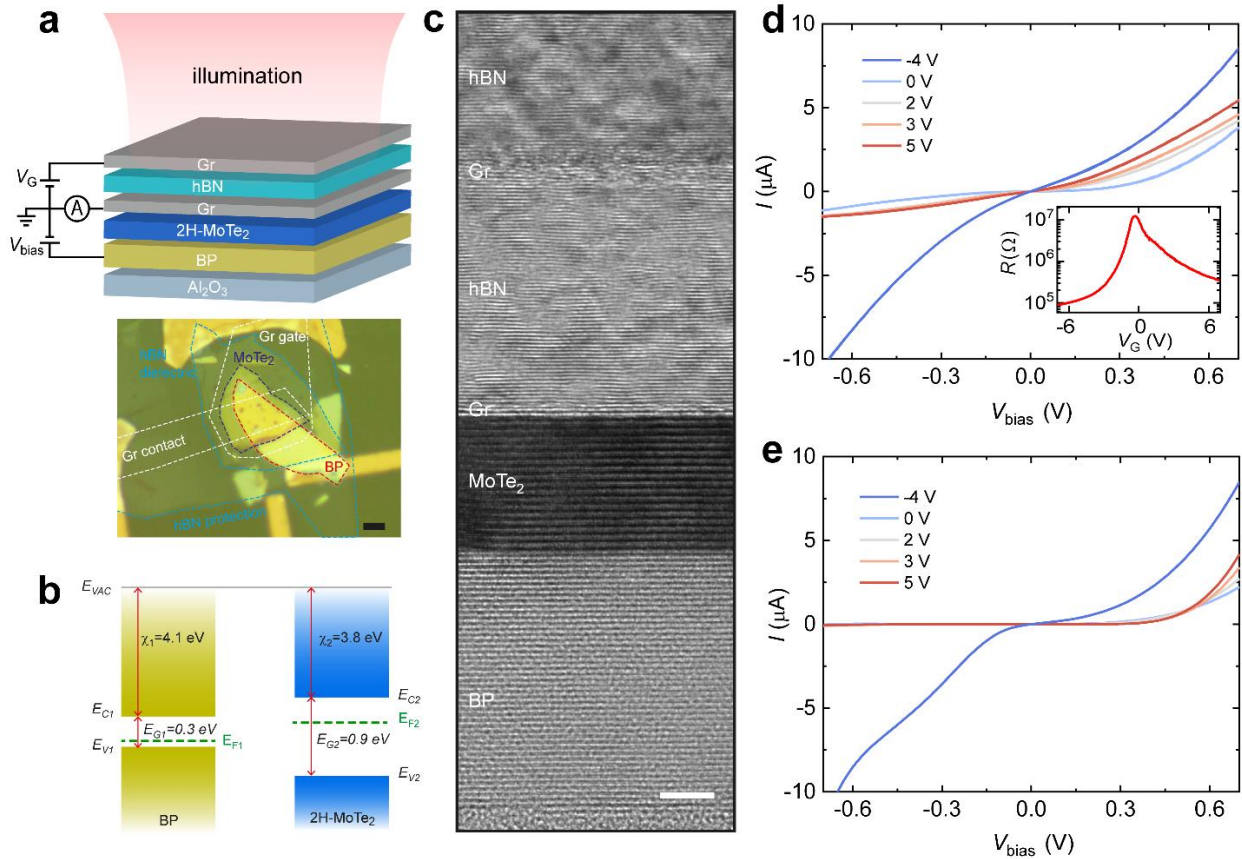

**Figure 1 | Design and working principle of a gate-tunable van der Waals photodetector. a,** (Top) Three-dimensional schematic of the device heterostructure and measurement scheme. The top protective hBN layer is not shown. (Bottom) Optical microscopy image of the actual device. Scale bar: 5 µm. **b,** Energy bands of BP and 2H-MoTe$_2$ before contact. **c,** Cross-sectional transmission electron microscopy image of the same heterostructure taken after device measurements. Scale bar: 5 nm. **d,** Measured dark current-voltage characteristics at different gate voltages. Inset: Low-bias (5 mV) device resistance vs. gate voltage. **e,** Simulated dark current-voltage characteristics at different gate voltages.

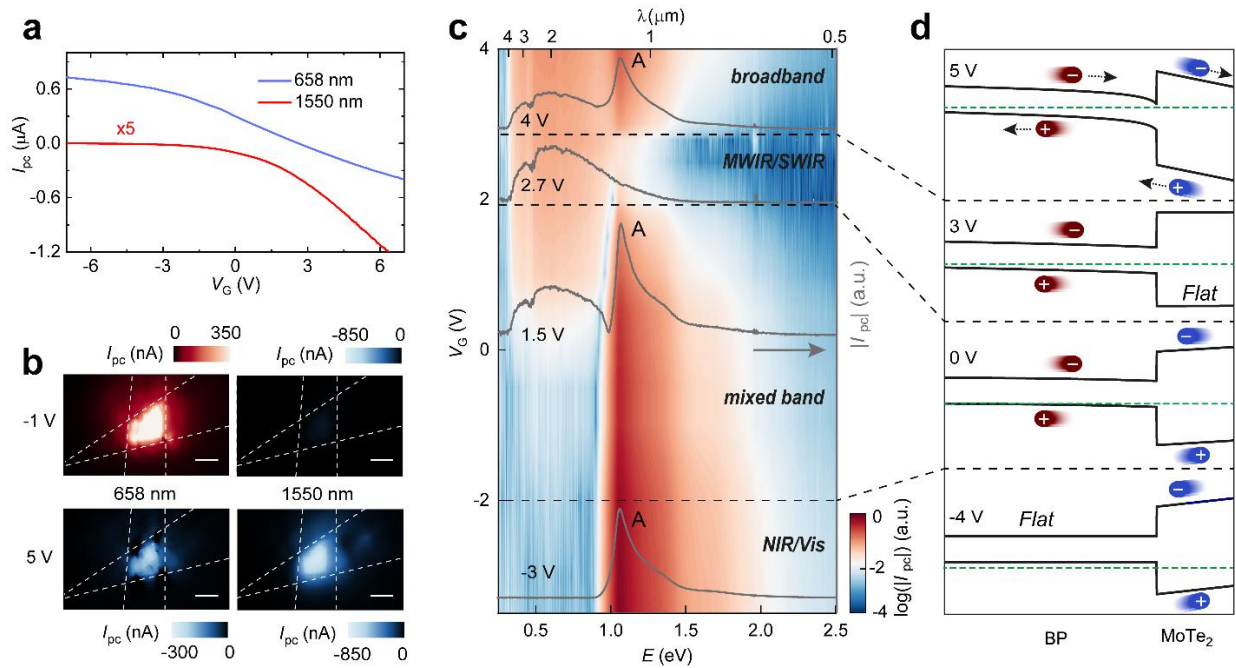

**Figure 2 | Gate-tunable spectral photoresponse. a,** zero-bias photocurrent $I_{pc}$ vs gate voltage $V_G$ for λ = 658 nm (25 µW) and 1550 nm (5 µW) excitation. **b,** Scanning photocurrent images of detector taken with λ = 658 nm (left) and 1550 nm (right) at $V_G$ = −1 V (top) and $V_G$ = 5 V (bottom). Scale bars: 5 µm. **c,** Continuous 2D plot of photocurrent spectra vs light energy and gate voltage. The spectrally dependent gate regions are divided by dashed lines into MWIR/SWIR band, NIR/Vis band, mixed band, and broadband. A representative photocurrent spectrum for each region is overlaid in gray. The peak at 1.06 eV corresponds to the A exciton of 2H-MoTe$_2$. **d,** Band structure schematics of BP / 2H-MoTe$_2$ junction under illumination from finite-element simulations for the representative gate voltages in **c**. The MoTe$_2$ and BP bands are flat at $V_G$ = 3 and $V_G$ = −4 V, respectively. Photocarriers generated by IR and visible light are colored in red and blue, respectively, with arrows and blurring denoting their directional movement caused by the local electric field.

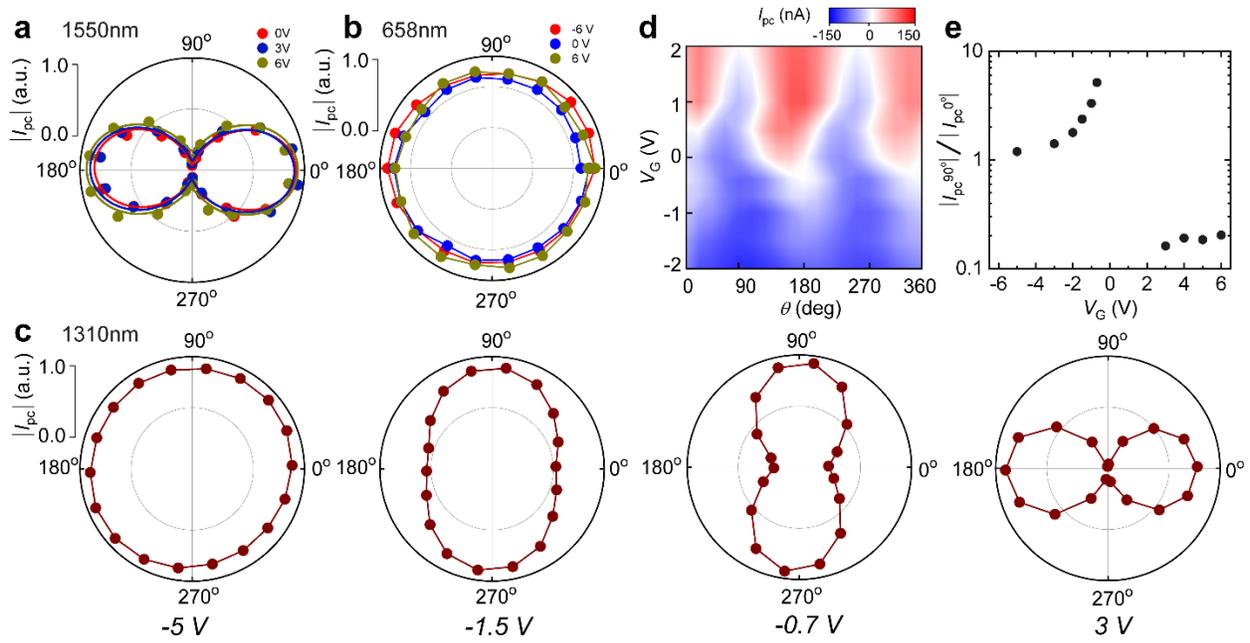

**Figure 3 | Gate-tunable polarization sensitivity.** $|I_{pc}|$ vs linear polarization angle at different gate voltages and with **a,** λ = 1550 nm, **b,** λ = 658 nm, and **c,** λ = 1310 nm. **d,** Continuous 2D plot of $I_{pc}$ vs linear polarization angle and gate voltage at λ = 1310 nm. **e,** Anisotropy ratio $|I_{pc}^{90°}|/|I_{pc}^{0°}|$ of gate-dependent photocurrent at λ = 1310 nm.

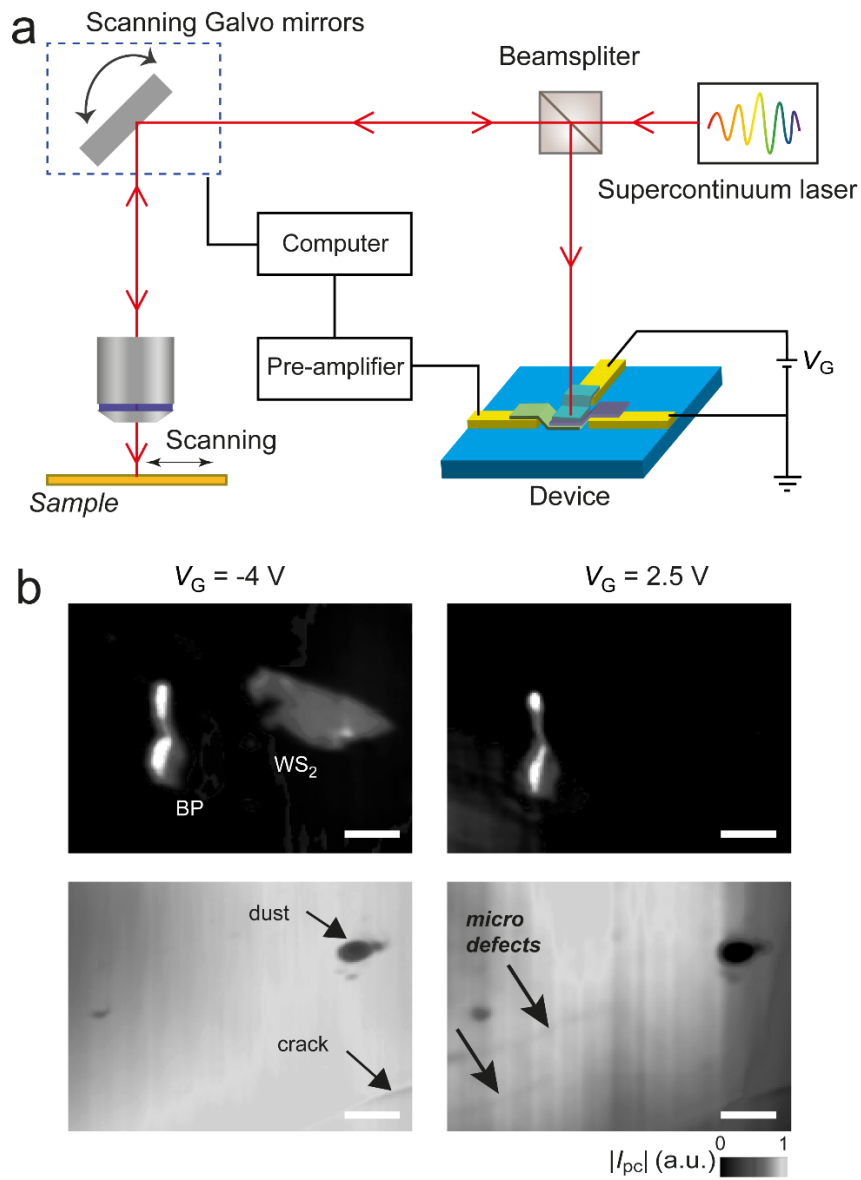

**Figure 4 | Spectrally tunable confocal imaging using van der Waals photodetector. a,** Schematic of imaging setup with scanning broadband laser and Gr / 2H-MoTe$_2$ / BP detector. **b,** Confocal reflection images of BP and WS$_2$ flakes on Al$_2$O$_3$ substrate (upper, scale bar = 20 µm) and InGaAs quantum well wafer (bottom, scale bar = 50 µm) taken under $V_G$ = −4 V (left, NIR/Vis band) and $V_G$ = 2.5 V (right, MWIR/SWIR band). Different features can be seen under the two imaging modes for each sample.

# Supporting Information

**Gate-Tunable Multi-Band van der Waals Photodetector and Polarization Sensor**

Daozhi Shen[1,2,3], HeeBong Yang[2,4], Tarun Patel[2,5], Daniel A. Rhodes[6], Thomas Timusk[7], Y. Norman Zhou[8,9], Na Young Kim[2,4,9], and Adam W. Tsen[2,3,9]*

[1] School of Mechanical Engineering, Shanghai Jiao Tong University, Shanghai, 200240, China
[2] Institute for Quantum Computing, University of Waterloo, Waterloo, ON, Canada N2L 3G1
[3] Department of Chemistry, University of Waterloo, Waterloo, ON, Canada N2L 3G1
[4] Department of Electrical and Computer Engineering, University of Waterloo, Waterloo, ON, Canada N2L 3G1
[5] Department of Physics and Astronomy, University of Waterloo, Waterloo, ON, Canada N2L 3G1
[6] Department of Materials Science and Engineering, University of Wisconsin-Madison, WI, USA
[7] Department of Physics and Astronomy, McMaster University, Hamilton, Ontario, Canada L8S 4M1
[8] Centre for Advanced Materials Joining and Department of Mechanical and Mechatronics Engineering, University of Waterloo, Waterloo, Ontario, Canada N2L 3G1
[9] Waterloo Institute for Nanotechnology, University of Waterloo, Waterloo, Ontario, Canada N2L 3G1

* Corresponding author: awtsen@uwaterloo.ca

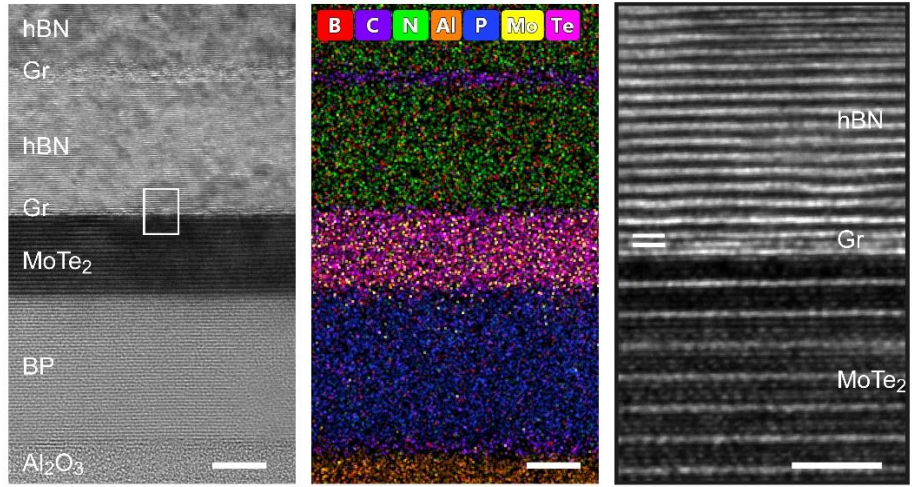

**Figure S1** (Left) Cross-sectional transmission electron microscopy image of the junction. Scale bar: 10 nm. (Middle) Corresponding energy dispersive spectroscopy image. Scale bar: 10 nm. (Right) High-resolution transmission electron microscopy image of the hBN/Gr/MoTe$_2$ interface. Scale bar: 2 nm.

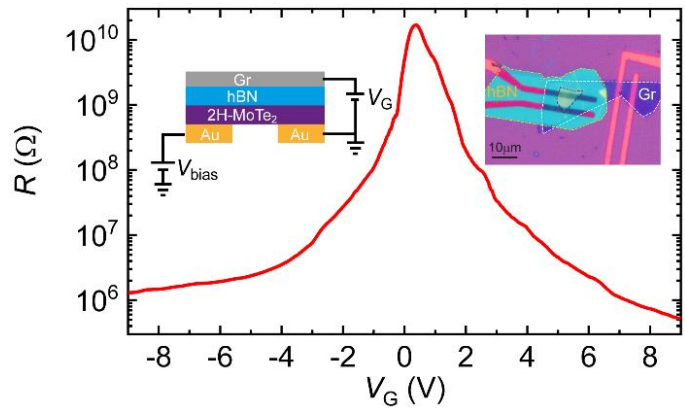

**Figure S2** Gate-dependent resistance of lateral 2H-MoTe$_2$ transistor. Insets show the measurement scheme and optical image of the device.

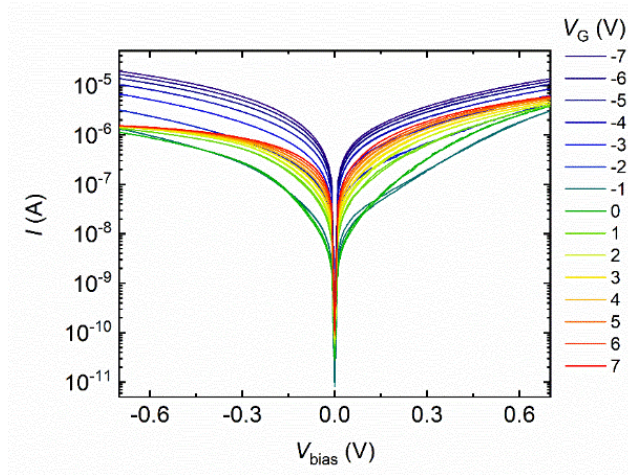

**Figure S3** Dark *IV* characteristics of the device under various gate voltages.

**Table S1** The values of hole doping concentration ($N_a$), electron doping concentration ($N_d$), and Fermi level ($E_F$) matched with gate voltage ($V_G$).

| | MoTe$_2$ | | | BP | | |
| --- | --- | --- | --- | --- | --- | --- |
| $V_G$ (V) | $N_a$ (1/cm$^3$) | $N_d$ (1/cm$^3$) | $E_F$ (eV) | $N_a$ (1/cm$^3$) | $N_d$ (1/cm$^3$) | $E_F$ (eV) |
| −4 | 8.00E+14 | - | -4.5 | 2.00E+18 | - | -4.38 |
| 0 | 1.00E+12 | - | -4.35 | 2.99E+17 | - | -4.33 |
| 2 | 3.00E+10 | - | -4.27 | 2.67E+16 | - | -4.27 |
| 3 | - | 6.20E+08 | -4.23 | - | 3.50E+15 | -4.25 |
| 5 | - | 1.20E+09 | -4.15 | - | 1.35E+17 | -4.19 |

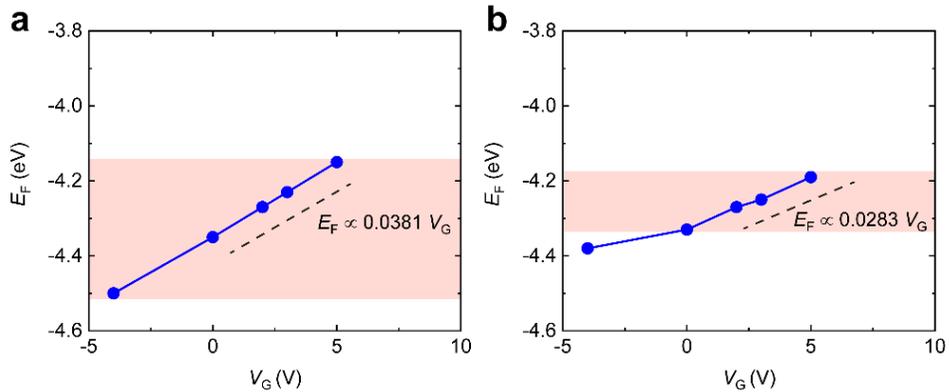

**Figure S4** Modeled $E_F$ vs $V_G$ for **a,** MoTe$_2$ and **b,** BP. The linear window is marked in pink. All the $E_F$ points in the simulation lie in the linear region of the $E_F$ - $V_G$ relation except for one outlier of $E_F = -4.38$ eV for BP. From the $E_F$ - $V_G$ relation, the gate-dependent band diagrams and *I-V* characteristics were subsequently obtained.

**Laser-based photocurrent measurements**

Laser diodes with wavelengths in the visible (λ = 658 nm, Thorlabs) and infrared (λ = 1310 and 1550 nm, Thorlabs) were focused on the detectors using a 15× reflective objective (Beck Optronic Solutions, NA = 0.5). A Glan-Taylor polarizer and a half-waveplate were used to control the linear polarization angle of the laser. A scanning mirror galvo in the beam path was used to both position the laser spot on the device (for fixed-point measurements) and raster-scan the laser across the device (for imaging), while the reflected light was monitored with a separate commercial photodetector (Thorlabs PDA36A and DET10D/M for the visible and infrared, respectively). Bias and gate voltages were applied to the detector using a Source Measure Unit (Keithley 2450 SMU) and the current was measured using a pre-amplifier (Stanford Research Systems SR570). Figure S5 shows the zero-bias photocurrent as a function of laser power for two wavelengths and various gate voltages. We have operated within the linear or nearly linear regime throughout our measurements. The full dark and illuminated *I-V* characteristics with various gate voltages are shown in Figure S3. While the 658 nm photocurrent response shows ambipolar behavior under zero bias, photocurrent for λ = 1550 nm is strongly activated only for positive $V_G$, which is consistent with the results in Figure 2a.

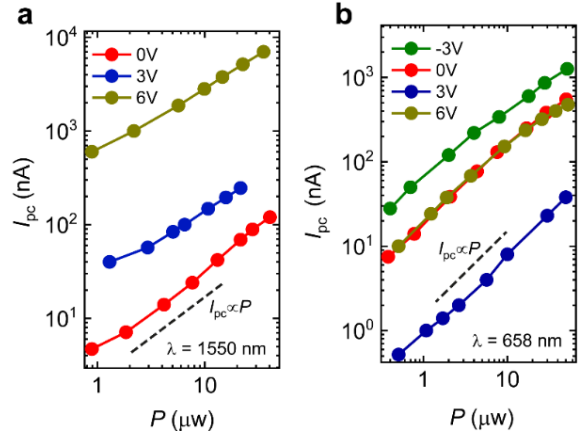

**Figure S5** Photocurrent $I_{pc}$ as function of laser power and gate voltage at λ of **a,** 1550 nm and **b,** 658 nm.

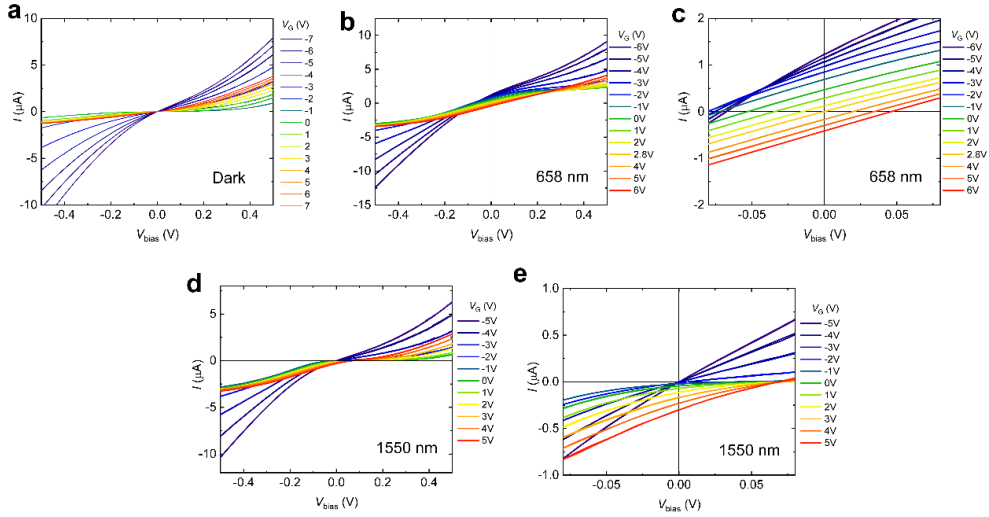

**Figure S6** Measured *I-V* characteristics of the device at different gate voltages under (a) dark environment, (b), (c) λ = 658 nm and (d), (e) 1550 nm laser excitation.

**Photocurrent spectroscopy**

Photocurrent spectra of the detectors were characterized using a Bruker IFS 66v/S Fourier transform infrared spectrometer (FTIR). Different biases and gate voltages were applied to the device using Source Meter Units (Keithley 2450). Illumination from a broadband tungsten halogen source was sent through the FTIR beamsplitter and then focused onto a ~ 3 mm spot centered on the device using a parabolic mirror (intensity ~ 2 W/cm$^2$). The generated photocurrent signal was sent through a current pre-amplifier (Stanford Research Systems SR570), a voltage pre-amplifier (Stanford Research Systems SR560), and then fed back into the FTIR, which Fourier-transforms the signal to determine the spectral response. Figure S7 shows that broadband photodetection can be achieved by applying a proper bias voltage; however, accessing each band individually can only be realized by gate tuning.

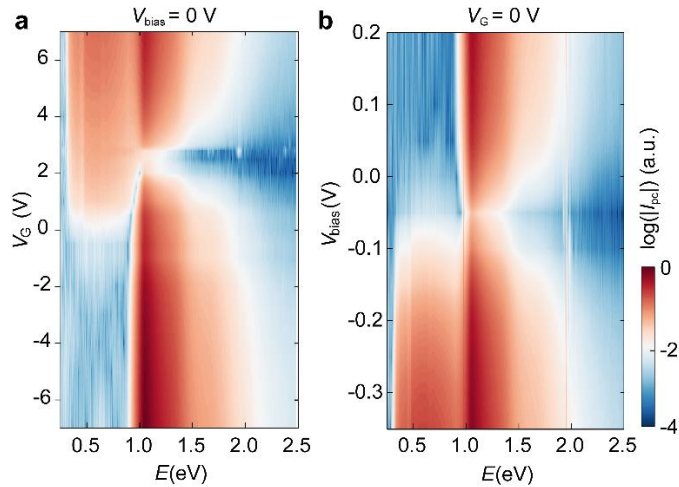

**Figure S7** Continuous 2D plots of photocurrent spectra vs light energy and **a,** gate voltage ($V_{bias}$ = 0 V) **b,** bias voltage ($V_G$ = 0 V).

**Responsivity, quantum efficiency, and detectivity characterization**

We first characterized the gate-dependent spectral responsivity ($\Re$) of the detector using the FTIR. The source intensity at different wavelengths was measured using a previously calibrated van der Waals photodetector from our prior work[2]. The spectral response of the photocurrent at various gate voltages for the device under test was then normalized to the spectrum of the source to obtain $\Re$ (see Figure S8, top). From $\Re$, we can determine the external quantum efficiency by EQE = $\Re hc/(\lambda e)$, where $h$ is Planck's constant, $c$ is the speed of light in vacuum, $\lambda$ is the wavelength, and $e$ is the electron charge (see Figure S8, bottom).

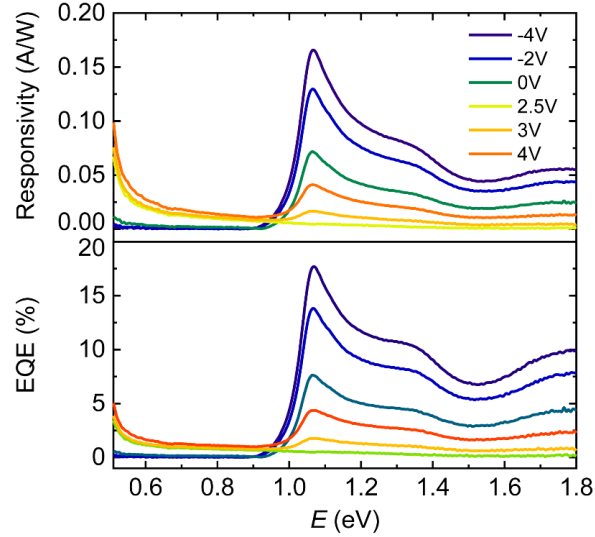

**Figure S8** Photodetector performance metrics. (Top) responsivity, (bottom) EQE of photodetector under varies gate voltages at zero bias.

Afterwards, we determined the specific detectivity $D^*$ of the device using two independent methods. First, $D^*$ was extracted from a noise measurement by enclosing the device in a dark, shielded environment under the specific voltage condition $V_G = 6$ V (broadband mode) and $V_{bias} = 0$ V. A lock-in amplifier (Stanford Research Systems SR830) was used to measure the current noise spectral density $i$ of the device and circuit as a function of frequency from 10 Hz to 100 kHz. Details can be found from our previous work [1]. Figure S9 shows current noise $i$ vs frequency contributed by the detector at $V_G = 6$ V and $V_{bias} = 0$ V. The specific detectivity can then be obtained from the relation $D^* = \Re A^{1/2}/S_n$, where $A$ is the detector area, and $S_n$ is the current noise spectral density extrapolated up to and averaged over the full electrical bandwidth (1 MHz). In this way, we obtain $D^* = 0.9 \times 10^9$ cm Hz$^{1/2}$ W$^{-1}$ at 0.52 eV (for $V_G = 6$ V and $V_{bias} = 0$ V). Second, we extracted $D^*$ another way using the formula[2-4]: $D^* = \Re(A\Delta f)^{1/2}/i_n = \Re(A\Delta f)^{1/2}/(4kT\Delta f/R + 2eI_{dark}\Delta f)^{1/2} = \Re A^{1/2}(4kT/R + 2eI_{dark})^{-1/2}$, where $i_n$ is the noise current, $k$ is Boltzmann constant, $T$ is the temperature, $R$ is the device resistance, $e$ is the electron charge, and $I_{dark}$ is the dark current. Figure S3 shows that the dark current of the device near $V_{bias} = 0$ is in the $10^{-11} \sim 10^{-10}$ A level. Under this condition, $4kT/R \gg 2eI_{dark}$ so $D^* \approx \Re A^{1/2}(4kT/R)^{-1/2}$. At 0.52 eV (and for $V_G = 6$ V and $V_{bias} = 0$ V), $D^*$ is calculated to be $1.1 \times 10^9$ cm Hz$^{1/2}$ W$^{-1}$, which is very close to the value obtained through the noise measurement. We therefore use the formula $D^* \approx \Re A^{1/2}(4kT/R)^{-1/2}$ to extract the full $D^*$ spectrum at various gate voltages (see Figure S10).

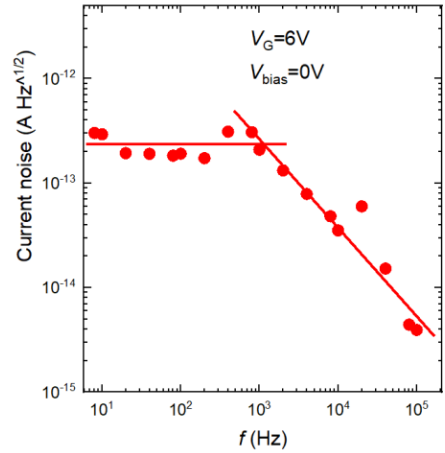

**Figure S9** Spectral noise density of the device at $V_G = 6$ V and $V_{bias} = 0$ V. The background noise has been subtracted and the data are fit to power laws.

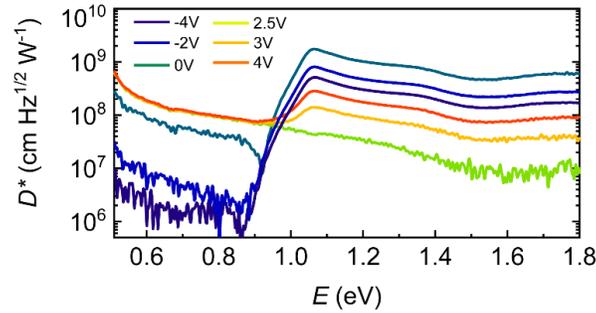

**Figure S10** $D^*$ of photodetector under various gate voltages with zero bias.

### Frequency response characterization

We measured the frequency response of our detector to obtain the −3 dB electrical bandwidth. The intensity of 658 nm ($MoTe_2$ response) and 1550 nm (BP response) wavelength laser diodes was modulated using a sinusoidal waveform with frequency ranging from 50 kHz to 2 MHz by a function generator (Stanford Research Systems DS345). The photocurrent was sent through a current pre-amplifier (Femto DHPCA-100) in high-bandwidth mode to an oscilloscope (Siglent, SDS1104X-E). We normalized the photocurrent signal at each frequency to the photoresponse from a Thorlabs PDA-36A detector (10 MHz bandwidth setting). The extracted bandwidth is relatively independent of the laser wavelength and gate voltage, as can be seen from Figure S11 below.

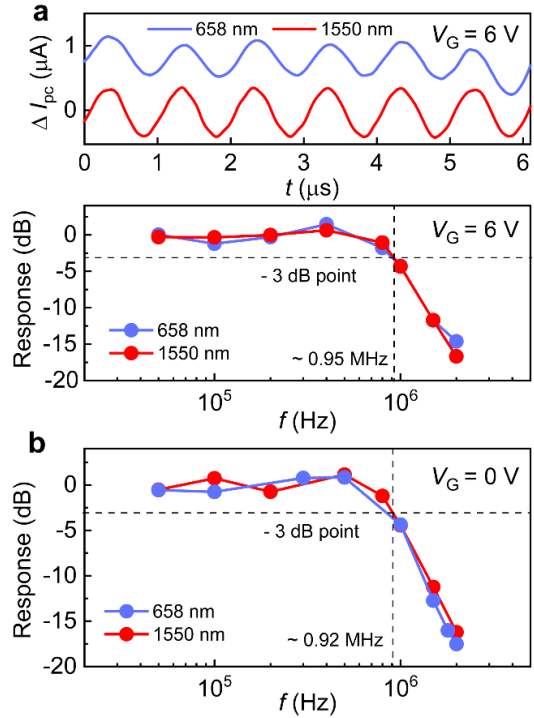

**Figure S11 a**, Photoresponse of the detector ($V_G$ = 6 V) under laser irradiation (λ = 658 nm and 1500 nm) intensity-modulated by a sinusoidal waveform at 1 MHz frequency. **b**, $I_{pc}$ frequency response for both wavelengths at $V_G$ = 6 V (upper) and $V_G$ = 0 V (lower).

| Materials | Bias voltage | Detectivity | Response time | Polarization sensitivity | Reference |
|---|---|---|---|---|---|
| BP/MoS$_2$/Si | 0 V and -0.5 V | $6.4 \times 10^9$ | 50 μs | No | Ref. 5 |
| PbS/MoS$_2$/BP | 0 V and 0.3 V | $5 \times 10^9$ | 20 μs | No | Ref. 6 |
| BP/MoS$_2$/MoTe$_2$ | -1 V and 1 V | - | 35 ns | No | Ref. 7 |
| Ge/MoS$_2$ | -0.5 V and -3.5 V | - | 10 ms | No | Ref. 8 |
| Gr/MoTe$_2$/BP | 0 | $2 \times 10^9$ | ~ 300 ns | Yes | This work |

**Table S1** Device performance comparison of the device with reported multiband or dual-band detectors based on 2D materials

**Spectral crosstalk characterization**

The crosstalk of the device can be extracted from the spectral response[5]. As the device operates in the MWIR/SWIR and NIR/Vis bands when the gate voltage is set to 2.7 V and -1 V, respectively, we calculate the crosstalk between those two bands from the photocurrent spectrum at these two gate voltages. As shown in Figure S12, $S_A$ and $S_B$ are the areas obtained by integrating the MWIR/SWIR and NIR/Vis response curves over the photon energy axis, respectively. S'$_A$ is the area of MWIR/SWIR response overlapped with the NIR/Vis band, and S'$_B$ is the area of NIR/Vis response overlapped with the MWIR/SWIR band. The crosstalk of the MWIR/SWIR band to NIR/Vis band is:

$$C_{\text{A to B}} = \frac{S'_A}{S_B} \times 100\% \sim 7\%,$$

and the crosstalk of the NIR/Vis band to MWIR/SWIR band is:

$$C_{\text{B to A}} = \frac{S'_B}{S_A} \times 100\% \sim 5\%.$$

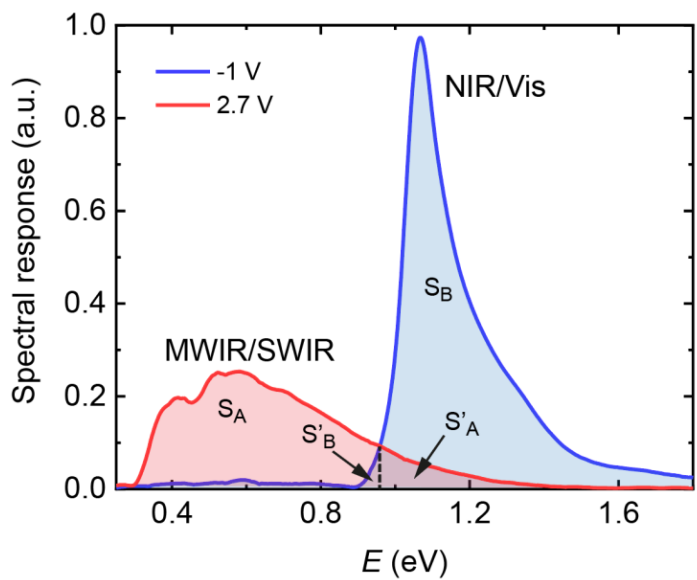

**Figure S12** Spectral crosstalk characterization of the device between the MWIR/SWIR and NIR/Vis bands.